\documentclass{phb-proc4-auth}

\usepackage{graphicx}
\usepackage{amssymb}

\begin{document}
\begin{frontmatter}


\journal{SCES '04}


\title{Resonant electronic Raman scattering near a quantum critical point}

%
%
%
%
%
%

\author[UA]{A.~M.~Shvaika\corauthref{1}}
\author[UA]{O.~Vorobyov}
\author[US]{J.~K.~Freericks}
\author[CA]{T.~P.~Devereaux}

%

\address[UA]{Institute for Condensed Matter Physics, Lviv, Ukraine} 
\address[US]{Department of Physics, Georgetown University, Washington, DC, USA}
\address[CA]{Department of Physics, University of Waterloo, Ontario, Canada}

%
%
%
%


%
%
%
%

\corauth[1]{Corresponding Author: Institute for Condensed Matter Physics of 
the National
Academy of Sciences of Ukraine, 1 Svientsitskii Street, 79011 Lviv, Ukraine,
Phone: +380-322-761054, Fax: +380-322-761158,
Email: ashv@icmp.lviv.ua}


\begin{abstract}
We calculate the resonant electronic Raman scattering for the
Falicov-Kimball model near the Mott transition on a hypercubic
lattice.  The solution is exact, and employs dynamical mean field theory.
\end{abstract}

%
%



\end{frontmatter}

%
%
%
%
%
The Falicov-Kimball model~\cite{falicov_kimball} has two kinds of particles:  
conduction electrons, which are mobile 
and localized electrons which are immobile.  
The Hamiltonian is (at half filling)
\begin{eqnarray}
\mathcal{H}&=&-\frac{t^*}{2\sqrt{d}}\sum_{\langle ij \rangle} (c^\dagger_ic_j
+c^\dagger_jc_i)\nonumber
\\
&&+U\sum_i \left(c^\dagger_ic_i-\frac12\right)\left(f^\dagger_if_i-\frac12\right),
\label{eq: ham_def}
\end{eqnarray}
where $c^\dagger_i$ ($c_i$) creates (destroys) a conduction electron at
site $i$, $f^\dagger_i$ ($f_i$) creates (destroys) a localized electron at
site $i$, $U$ is the on-site Coulomb interaction between the
electrons, and $t^*$ is the hopping integral~\cite{metzner_vollhardt}
(which we use as our energy unit).
The symbol $d$ is the spatial dimension, and
$\langle i j\rangle$ denotes a sum over all nearest neighbor pairs
(we work on a hypercubic lattice).
The model is exactly solvable with dynamical mean field 
theory~\cite{brandt_mielsch} when $d\rightarrow\infty$
(see ~\cite{freericks_review} for a review).

The formalism for calculating the Raman response was originally developed by
Shastry and Shraiman~\cite{shastry_shraiman}. The expression for the 
Raman response is
\begin{eqnarray}
R(\Omega)&=&2\pi \sum_{i,f} \exp(-\beta\varepsilon_i)
    \delta(\varepsilon_f - \varepsilon_i - \Omega)\nonumber
    \\
&\times& \left| \frac{hc^2}{V\sqrt{\omega_i\omega_o}}e_\alpha^i e_\beta^o
    \left\langle f \left| \hat M^{\alpha\beta}\right| i \right\rangle
    \right|^2 /\mathcal{Z}
\label{eq: raman1}
\end{eqnarray}
for the scattering of electrons by optical photons
(the repeated indices $\alpha$ and $\beta$ are summed over).
Here ${\bf e}^{i(o)}$ denotes the polarization vector of the incident and 
outgoing photon, $\varepsilon_{i(f)}$
refer to the eigenstates describing the ``electronic matter'', and
$\mathcal{Z}$ is the partition function.  The coupling of the photon
to the electronic system is treated with the linear coupling of the vector
potential to the current operator ${\bf j}_\alpha=\sum_{\bf k}
\partial\varepsilon(\bf k)/
\partial k_\alpha c^\dagger_{\bf k}c_{\bf k}$, and the quadratic coupling of the
vector potential to the stress-tensor operator $\gamma_{\alpha\beta}=
\sum_{\bf k}\partial^2\varepsilon(\bf k)/\partial k_\alpha\partial k_\beta
c^\dagger_{\bf k}c_{\bf k}$, with $\varepsilon({\bf k})$ the bandstructure
and $c_{\bf k}$ the destruction operator for an electron with momentum 
${\bf k}$.  
The scattering operator then becomes
\begin{eqnarray}\label{M_oper}
    \left\langle f \left| \hat M^{\alpha\beta} \right| i \right\rangle
    &=&
    \left\langle f \left| \gamma_{\alpha,\beta} \right| i \right\rangle\nonumber
    \\
    &+& \sum_l \left(
    \frac{\left\langle f \left| j_{\beta} \right| l \right\rangle
    \left\langle l \left| j_{\alpha} \right| i \right\rangle}
    {\varepsilon_l - \varepsilon_i - \omega_i}
    \right.\nonumber
    \\
    &+&
    \left.
    \frac{\left\langle f \left| j_{\alpha} \right| l \right\rangle
    \left\langle l \left| j_{\beta} \right| i \right\rangle}
    {\varepsilon_l - \varepsilon_i + \omega_o}
    \right),
\end{eqnarray}
with the sum $l$ over intermediate states.

We have evaluated these 
expressions for the Stokes Raman response, with an incident photon
frequency $\omega_i$, an outgoing photon frequency $\omega_o$, and a
transfered photon frequency $\Omega=\omega_i-\omega_o$.  The procedure is
complicated, and involves first computing the response functions on the 
imaginary time axis, then Fourier transforming to imaginary frequencies,
and finally performing an analytic continuation to the real axis~\cite{details}.
Our calculations
include effects from nonresonant diagrams, from resonant diagrams, and
from so-called mixed diagrams.  We analyze three different symmetries of the
incident and outgoing light: $A_{\textrm{1g}}$ with both incident and
outgoing polarizers aligned along the diagonal ${\bf e}^i={\bf e}^o=
(1,1,...)$, $B_{\textrm{1g}}$
with the incident light polarized
along one diagonal ${\bf e}^i=(1,1,...)$, and the outgoing light along 
another diagonal ${\bf e}^o=(1,-1,1,-1,...)$, and $B_{\textrm{2g}}$ with the 
incident light polarized along ${\bf e}^i=(1,0,1,0,1,0,...)$ 
and the outgoing light polarized along
${\bf e}^o=(0,1,0,1,0,1,...)$.  The $A_{\textrm{1g}}$ response has contributions
from all types of processes, the $B_{\textrm{1g}}$ response is nonresonant or
resonant, and the $B_{\textrm{2g}}$ response is purely resonant.

The Falicov-Kimball model on a hypercubic lattice has a Mott transition
at half filling when $U=\sqrt{2}$.  The transition is to a pseudogap-like
phase, because the infinite tails of the noninteracting density of states (DOS)
(which is a Gaussian) do not allow the system to have a true gap of finite 
width.  Instead the DOS is equal to zero only at the chemical potential,
and there is exponentially small DOS within a ``gap region''. We examine the
system just on the insulating side of the Mott transition at $U=1.5$.

\begin{figure}[htbf]
\centering
\includegraphics[width=2.55in]{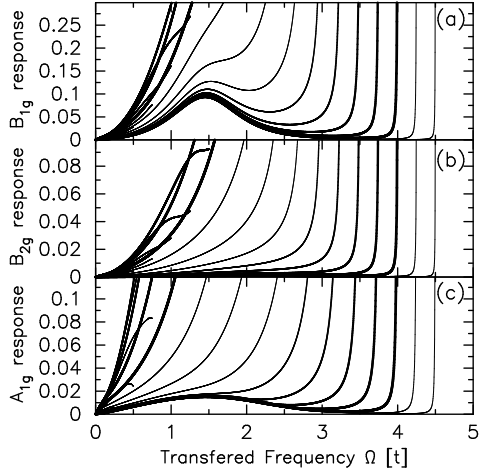}
\caption{The (Stokes) resonant Raman response for
three different symmetries as a function of $\Omega$ (at $T=0.05$ and
$U=1.5$; just on the insulating side of the Mott transition). The curves 
are for 
different values of the incident photon frequency.  \label{fig: raman}}
\end{figure}

\begin{figure}
\centering
\includegraphics[width=2.55in]{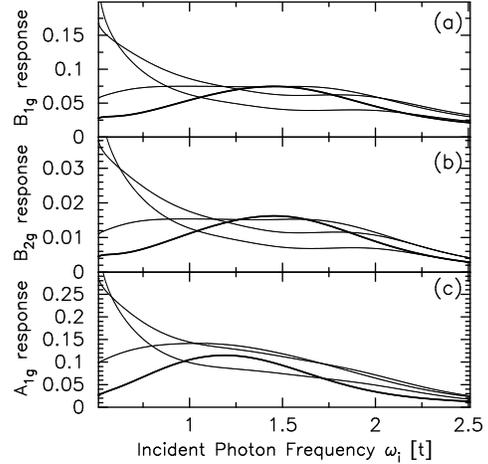}
\caption{ Resonant profile for $\Omega=0.5$ and various $\omega_i>0.5$.
The curves correspond to $T=1$, $0.5$, $0.2$, and $0.05$ in order of thinnest 
to thickest.
\label{fig: raman2}}
\end{figure}

Our results for the Raman response as a function of the transfered frequency
$\Omega$ appear in Fig.~\ref{fig: raman}.  Note how there are
strong resonant effects, including the sharp peak of the triple resonance
($\Omega=\omega_i$).
Note further that the full response is not just an enhancement of the 
nonresonant features (which are apparent when the incident photon frequency
becomes large), but that the shape of the response can change dramatically
due to resonant effects. This is most apparent when the incident photon
energy is close to $U$. In Fig.~\ref{fig: raman2}, we plot the resonant
profile at fixed $\Omega=0.5$ as a function of the incident photon frequency
$\omega_i$.  Note how the low energy features change
their resonant behavior from being centered around
$\omega_i\approx 0.5=\Omega$ (dominated by 
triple-resonance effects) at high temperature to being centered around
$\omega_i\approx U$ at lower temperatures. This resonance of a low-energy 
feature, due to a higher-energy photon energy has been often seen in
Raman scattering in strongly-correlated materials.

\textit{Acknowledgments}: We 
acknowledge support from the CRDF 
(UP2-2436-LV-02), from the NSF (DMR-0210717) (J.K.F.)
and from NSERC, PREA, and the Alexander von Humboldt foundation (T.P.D.). 
%
%
%
%

%
%
%
%



\begin{thebibliography}{00}

\bibitem{falicov_kimball}
L.~M.~Falicov and J.~C.~Kimball, Phys. Rev. Lett. {\bf 22} (1969) 997.
\bibitem{metzner_vollhardt}
W.~Metzner and D.~Vollhardt, Phys. Rev. Lett. {\bf 62} (1989) 324.
\bibitem{brandt_mielsch}
U.~Brandt and C.~Mielsch, Z. Phys. B---Condens. Mat. {\bf 75} (1989) 365;
{\bf 79} (1990) 295; {\bf 82}, (1991) 37.
\bibitem{freericks_review}
J.~K.~Freericks and V.~Zlati\'c, Rev. Mod. Phys. {\bf 75} (2003) 1333.
\bibitem{shastry_shraiman}
B.~S.~Shastry and B.~I.~Shraiman, Phys. Rev. Lett. {\bf 65} (1990) 1068; 
Int. J. Mod. Phys. B {\bf 5}, (1991) 365.
\bibitem{details} A.~M.~Shvaika, O.~Vorobyov, J.~K.~Freericks, and 
T.~P.~Devereaux, unpublished.
\end{thebibliography}
\end{document}